\documentclass[preprint,aps,prd,nofootinbib,superscriptaddress]{revtex4}
\usepackage{hyperref}
\usepackage[pdftex]{graphicx,color}
\usepackage{slashed,simplewick}
\usepackage{amsmath,amssymb}
\usepackage{hhline}
\usepackage{subfigure}
\usepackage{epstopdf}
\usepackage{epsfig}
\usepackage{graphicx}
\usepackage{caption}
\usepackage{nicefrac,xfrac}

\newcommand{\ket}[1]{\left|#1\right\rangle}
\newcommand{\bra}[1]{\left\langle#1\right|}
\newcommand{\braket}[2]{\left\langle #1 | #2 \right\rangle}

\newcommand{\beq}{\begin{equation}}
\newcommand{\eeq}{\end{equation}}
\newcommand{\beqa}{\begin{eqnarray}}
\newcommand{\eeqa}{\end{eqnarray}}

\newcommand{\mmbar}{M_\mu-\overline{M}_\mu}
\newcommand{\mmu}{{M_\mu}}
\newcommand{\mmup}{{M_\mu^P}}
\newcommand{\mmuv}{{M_\mu^V}}
\newcommand{\ammu}{{\overline{M}_\mu}}
\newcommand{\ammup}{{\bar{M}_\mu^P}}
\newcommand{\ammuv}{{\bar{M}_\mu^V}}

\begin{document}
\begin{flushright}
WSU-HEP-2002\\

\end{flushright}
\title{\boldmath Muonium-antimuonium oscillations in effective field theory}

\author{Renae Conlin}
\affiliation{Department of Physics and Astronomy\\
        Wayne State University, Detroit, MI 48201, USA}

\author{Alexey A.\ Petrov}
\affiliation{Department of Physics and Astronomy\\
        Wayne State University, Detroit, MI 48201, USA}
\affiliation{Leinweber Center for Theoretical Physics\\
        University of Michigan, Ann Arbor, MI 48196, USA}


\begin{abstract}
\noindent
Flavor violating processes in the lepton sector have highly suppressed branching ratios in the 
standard model mainly due to the tiny neutrino mass. This means that observing lepton flavor violation (LFV)
in the next round of experiments would constitute a clear indication of physics beyond the standard model (BSM). 
We revisit a discussion of one possible way to search for LFV, muonium-antimuonium oscillations. 
This process violates muon lepton number by two units and could be sensitive to the types of BSM physics that 
are not probed by other types of LFV processes. Using techniques of effective field theory, we calculate the 
mass and width differences of the mass eigenstates of muonium. We argue that its invisible decays give the 
parametrically leading contribution to the lifetime difference and put constraints on the scales of new physics 
probed by effective operators in muonium oscillations.
\end{abstract}

\maketitle

\section{Introduction}\label{Intro}

Flavor-changing neutral current (FCNC) interactions serve as a powerful probe of physics beyond the standard 
model (BSM). Since no local operators generate FCNCs in the standard model (SM) at tree level, new physics (NP) degrees of 
freedom can effectively compete with the SM particles running in the loop graphs, making their discovery possible.
This is, of course, only true provided the BSM models include flavor-violating interactions. 

An especially clean system to study BSM effects in lepton sector is muonium $\mmu$, a QED bound state of 
a positively-charged muon and a negatively-charged electron, $|\mmu\rangle \equiv |\mu^+e^-\rangle$. 
The main decay channel for both states is driven by the weak decay of the muon. The average lifetime of a muonium state $\tau_\mmu$ is 
expected to be the same as that of the muon, $\tau_\mu = (2.1969811 \pm 0.0000022)\times 10^{-6}$ s \cite{Tanabashi:2018oca},
apart from the tiny effect due to time dilation, 
$(\tau_\mmu - \tau_\mu)/\tau_\mu = \alpha^2m_e^2/(2 m_\mu^2) = 6\times 10^{-10}$ \cite{Czarnecki:1999yj}.
Just like a positronium or a Hydrogen atom, muonium could be produced in two spin configurations, a spin-one triplet state 
called {\it ortho-muonium}, and a spin-zero singlet state called {\it para-muonium}. 
We shall denote the para-muonium state as $\ket{\mmup}$ and the ortho-muonium state as $\ket{\mmuv}$. 
If the spin of the state does not matter, we shall employ the notation $\ket{\mmu}$.

So far, we have not yet observed FCNC in the charged lepton sector. This is because in the standard model with massive neutrinos the 
charged lepton flavor violating (CLFV) transitions are suppressed by the powers of $m^2_\nu/m^2_W$, which renders the predictions for 
their transition rates vanishingly small, e.g. ${\cal B}  (\mu \to e\gamma)_{\nu SM} \sim 10^{-54}$  \cite{Raidal:2008jk,Bernstein:2013hba}. 
Yet, experimental analyses constantly push the bounds on the CLFV transitions. It might be that in some models of NP, such as a model with 
the doubly-charged Higgs particles \cite{Swartz:1989qz,Chang:1989uk,Kiers:2005gh,Kiers:2005vx}, the effective $\Delta L=2$ transitions could occur at a rate 
that is not far below the sensitivity of currently-operating experiments. Alternatively, it might 
be that no term that changes the lepton flavor by two units is present in a BSM Lagrangian. But even in this case, a subsequent application of
two  $\Delta L=1$ interactions would also generate an {\it effective} $\Delta L=2$ interaction.
 
Such a $\Delta L=2$ interaction would then change the muonium state into the anti-muonium one, leading to the possibility of 
muonium-anti-muonium oscillations. As a variety of well-established new physics models contain $\Delta L=2$ interaction
terms \cite{Raidal:2008jk}, observation of muonium converting  into anti-muonium could then provide especially clean probes 
of new physics in the leptonic sector \cite{Bernstein:2013hba,Willmann:1998gd}. Theoretical analyses of conversion probability 
for such transitions have been actively studied, mainly using the framework of particular 
models \cite{Pontecorvo:1957cp,Feinberg:1961zza,ClarkLove:2004,CDKK:2005,Li:2019xvv,Endo:2020mev}.
It would be useful to perform a model-independent computation of the oscillation parameters using techniques of effective theory that includes 
all possible BSM models encoded in a few Wilson coefficients of effective operators. We do so in this paper, computing all
relevant QED matrix elements. 
Finally, employing similar effective field theory (EFT) techniques for computation of the contributions that are non-local at the
muon mass scale, we present them in terms of the series of local operators expanded in 
inverse powers of $m_\mu$ \cite{Beneke:1996gn,Golowich:2005pt}. 

In this paper we discuss the most general analysis of $\mmbar$ oscillations in the framework of effective field theory. We review 
phenomenology of muonium oscillations in Sec. \ref{Pheno}, taking into account both mass and lifetime differences in the  
muonium system. We compute the mass and width differences in Sec. \ref{Oscillations}. In Sec. \ref{Constraints} we constrain the 
BSM scale $\Lambda$ using experimental muonium-anti-muonium oscillation parameters. We conclude in Sec. \ref{Conclusions}.
Appendix \ref{Appendix} contains some details of calculations.

\section{Phenomenology of Muonium Oscillations}\label{Pheno}

Phenomenology of $\mmbar$ oscillations is very similar to phenomenology of meson-antimeson oscillations \cite{Donoghue,Nierste}. There are, however, several
important differences that we will emphasize below. One major difference is related to the fact that both ortho- and para-muonium can, 
in principle, oscillate. While most studies only considered muonium oscillations due to the BSM heavy states, below we also discuss the possibility 
of oscillations via the light states. Since such states can go on mass shell, these contributions would lead to the possibility of a lifetime difference in 
the $\mmbar$ system.
 
If the new physics Lagrangian includes lepton-flavor violating interactions, the time development of a muonium and anti-muonium states 
would be coupled, so it would be appropriate to consider their combined evolution, 
\beq
|\psi(t)\rangle = 
\left( {\begin{array}{c}
   a(t) \\
   b(t) \\
  \end{array} } \right) =
  a(t) |\mmu\rangle + b(t) |\overline\mmu\rangle.
\eeq
The time evolution of $|\psi(t)\rangle$ evolution is governed by a Schr\"odinger equation,  
\begin{equation}\label{two state time evolution}
i\frac{d}{dt}
\begin{pmatrix}
\ket{M_\mu(t)} \\ \ket{\overline{M}_\mu(t)}
\end{pmatrix}
=
\left(m-i\frac{\Gamma}{2}\right)
\begin{pmatrix}
\ket{M_\mu(t)} \\ \ket{\overline{M}_\mu(t)}
\end{pmatrix}.
\end{equation}
CPT-invariance dictates that the masses and widths of muonium and anti-muonium are 
the same, $m_{11}= m_{22}$, $\Gamma_{11}=\Gamma_{22}$, while CP-invariance of the $\Delta L_\mu = 2$ interaction, which we 
assume for simplicity, dictates that
\begin{eqnarray}\label{off_diagonal_elements}
m_{12}=m^{*}_{21}, \qquad \Gamma_{12}=\Gamma^{*}_{21}.
\end{eqnarray}
The presence of off-diagonal pieces in the mass matrix signals that it needs to be diagonalized. 
The mass eigenstates $ |\mmu_{1,2} \rangle$ can be defined as
\beq
 |\mmu_{1,2} \rangle = \frac{1}{\sqrt{2}} \left[ |\mmu \rangle \mp  |\ammu \rangle
 \right] ,
\eeq
where we neglected CP-violation and employed a convention where $CP |\mmu_\pm \rangle = \mp  |\mmu_\pm \rangle$.
The mass and the width differences of the mass eigenstates are 
\begin{eqnarray}\label{def of mass and width difference}
\Delta m \equiv M_{1}-M_{2}, \qquad \Delta \Gamma \equiv \Gamma_{2}-\Gamma_{1}.
\end{eqnarray} 
where  $M_i$ ($\Gamma_i$) are the masses (widths) of the mass eigenstates $ |\mmu_{1,2} \rangle$. We defined $\Delta m$
and $\Delta \Gamma$ to be either positive or negative, which is to be determined by experiment. 
It is often convenient to introduce dimensionless quantities,
\beq\label{XandY}
x = \frac{\Delta m}{\Gamma}, \qquad y =  \frac{\Delta \Gamma}{2\Gamma},
\eeq
where the average lifetime $\Gamma=(\Gamma_1+\Gamma_2)/2$. It is important to note that while $\Gamma$ is 
defined by the standard model decay rate of the muon, $x$ and $y$ are driven by the lepton-flavor violating interactions. 
It is then expected that both $x,y \ll 1$.

The time evolution of flavor eigenstates follows from Eq.~(\ref{two state time evolution}) \cite{Donoghue,Nierste},
\beqa
\ket{M(t)} &=& g_+(t) \ket{\mmu} + g_-(t) \ket{\ammu},
\nonumber \\
\ket{\overline{M}(t)} &=& g_-(t) \ket{\mmu} + g_+(t) \ket{\ammu},
\eeqa
where the coefficients $g_\pm(t)$ are defined as
\beq\label{TimeDep}
g_\pm(t) = \frac{1}{2} e^{-\Gamma_1 t/2} e^{-iM_1t} \left[1 \pm e^{\Delta\Gamma t/2} e^{i \Delta m t} \right].
\eeq
As $x,y \ll 1$ we can expand Eq.~(\ref{TimeDep}) to get
\beqa
g_+(t) &=& e^{-\Gamma_1 t/2} e^{-iM_1t} \left[1 + \frac{1}{8} \left(y-ix\right)^2 \left(\Gamma t\right)^2 \right],
\nonumber \\
g_-(t) &=& \frac{1}{2} e^{-\Gamma_1 t/2} e^{-iM_1t}  \left(y-ix\right) \left(\Gamma t\right).
\eeqa
Denoting an amplitude for the muonium decay into a final state $f$ as $A_f = \langle f|{\cal H} |\mmu\rangle$ and 
an amplitude for its decay into a CP-conjugated final state $\overline{f}$ as $A_{\bar f} = \langle \overline{f}|{\cal H} |\mmu\rangle$,
we can write the time-dependent decay rate of $\mmu$ into the $\overline{f}$,
\beq
\Gamma(\mmu \to \overline{f})(t) = \frac{1}{2} N_f \left|A_f\right|^2 e^{-\Gamma t} \left(\Gamma t\right)^2 R_M(x,y),
\eeq
where $N_f$ is a phase-space factor and $R_M(x,y)$ is the oscillation rate,
\beq
R_M(x,y) = \frac{1}{2} \left(x^2+y^2\right).
\eeq
Integrating over time and normalizing to $\Gamma(\mmu \to f)$ we get the probability of $\mmu$ decaying as $\ammu$ at some time $t > 0$,
\beq\label{Prob_osc}
P(\mmu \rightarrow \ammu) = \frac{\Gamma(\mmu \to \overline{f})}{\Gamma(\mmu \to f)} = R_M(x,y). 
\eeq
This equation generalizes oscillation probability computed in the classic papers \cite{Feinberg:1961zza,CDKK:2005}
by accounting for the lifetime difference in the muonium system, making it dependent on both the normalized mass $x$ and 
the lifetime $y$ differences. We will compute those in the next section.

We shall use the data from the most recent experiment \cite{Willmann:1998gd} in order to place constraints on the 
oscillation parameters. To do so, we have to account for the fact that the set-up described in \cite{Willmann:1998gd} had 
muonia propagating in a magnetic field $B_0$. This magnetic field suppresses oscillations by removing degeneracy 
between $\mmu$ and $\ammu$. It also has a different effect on different spin configurations of the muonium state 
and the Lorentz structure of the operators that generate mixing \cite{Cuypers:1996ia,Horikawa:1995ae}. 
Experimentally these effects were accounted for by introducing a factor $S_B(B_0)$. The oscillation probability is 
then \cite{Willmann:1998gd},
\beq\label{Exp_constr}
P(\mmu \rightarrow \ammu) \leq 8.3 \times 10^{-11}/S_B(B_0).
\eeq
We shall use different values of $S_B(B_0)$, presented in Table II of  \cite{Willmann:1998gd}
when placing constraints on the Wilson coefficients of effective operators in the next section.

\section{Effective Theory of Oscillations}\label{Oscillations}

Muonium-anti-muonium oscillations could be effective probes of flavor-violating new physics in leptons.
One of the issues is that at this point we do not know which particular model of new physics will provide the 
correct ultraviolet (UV) extension for the standard model. However, since the muonium mass is most likely 
much smaller than the new particle masses, it is not necessary to know it. Any new physics scenario which involves 
lepton flavor violating interactions can be matched to an effective Lagrangian, ${\cal L}_{\rm eff}$, whose 
Wilson coefficients would be determined by the UV physics that becomes active at some 
scale $\Lambda$ \cite{Grzadkowski:2010es,Petrov:2016azi}, 
\beq\label{lagrang}
{\cal L}_{\rm eff}=-\frac{1}{\Lambda^{2}}\underset{i}{\sum}c_{i}(\mu)Q_{i},
\eeq
where the $c_{i}$'s are the short distance Wilson coefficients. They encode all model-specific information. 
$Q_{i}$'s are the effective operators which reflect degrees of freedom relevant at the scale at which a given process 
takes place. If we assume that no new light particles (such as ``dark photons" or axions) exist in the 
low energy spectrum, those operators would be written entirely in terms of the SM degrees of freedom. 
In the case at hand, all SM particles with masses larger than that of the muon should also be integrated out, 
leaving only muon, electron, photon, and neutrino degrees of freedom. 

It would be convenient for us to classify effective operators in Eq.~(\ref{lagrang}) by their lepton quantum numbers. 
In particular, we can write the effective Lagrangian as
\beq\label{LeptNumbLagr}
{\cal L}_{\rm eff} = {\cal L}_{\rm eff}^{\Delta L_\mu=0} + {\cal L}_{\rm eff}^{\Delta L_\mu=1} + {\cal L}_{\rm eff}^{\Delta L_\mu=2}
\eeq
The first term in this expansion contains both the standard model and the new physics contributions. It then follows that the 
leading term in ${\cal L}_{\rm eff}^{\Delta L_\mu=0}$ is suppressed by powers of $M_W$, not the new physics scale $\Lambda$. 
We should emphasize that only the operators that are local at the scale of the muonium mass are retained in 
Eq.~(\ref{LeptNumbLagr}).

The second term contains $\Delta L_{\mu} = 1$ operators. As we integrated out all heavy degrees of freedom,
the operators of lowest possible dimension that governs muonium oscillations must be of dimension six. 
The most general dimension six effective Lagrangian, $\mathcal{L}_{\rm eff}^{\Delta L_{\mu}=1}$, has the 
form \cite{Celis:2014asa,Hazard:2017udp}
\begin{eqnarray}\label{L_eff1}
{\cal L}_{\rm eff}^{\Delta L_{\mu}=1} = &-&\frac{1}{\Lambda^2} \sum_f \Big[
\left( C_{VR}^{f} \ \overline\mu_R \gamma^\alpha e_R + 
C_{VL}^{f} \ \overline\mu_L \gamma^\alpha e_L \right) \ \overline f \gamma_\alpha f 
\nonumber \\
&+& \
\left( C_{AR}^{f} \ \overline\mu_R \gamma^\alpha e_R + 
C_{AL}^{q} \ \overline\mu_L \gamma^\alpha e_L \right) \ \overline f \gamma_\alpha \gamma_5 f 
\nonumber \\
&+& \
m_e m_f G_F \left( C_{SR}^{f} \ \overline\mu_R e_L + 
C_{SL}^{f} \ \overline\mu_L e_R \right) \ \overline f f 
\\
&+& \
m_e m_f G_F \left( C_{PR}^{f} \ \overline\mu_R e_L + 
C_{PL}^{f} \ \overline\mu_L e_R \right) \ \overline f \gamma_5 f
\nonumber \\
&+& \
m_e m_f G_F \left( C_{TR}^{f} \ \overline\mu_R \sigma^{\alpha\beta} e_L + 
C_{TL}^{f} \ \overline\mu_L \sigma^{\alpha\beta} e_R \right) \ \overline f \sigma_{\alpha\beta} f 
 + h.c. ~ \Big],
\nonumber
\end{eqnarray}
where $G_F \sim M_W^{-2}$ is the Fermi constant, $\mu$ and $e$ are the fermion fields, 
$(\mu,e)_{L,R} = P_{L,R}(\mu,e)$. $P_{R,L} = \frac{1}{2}\left(1\pm\gamma^{5}\right)$ are the projection 
operators, and $f$ represents other fermions that are not integrated out at the the muonium scale.
The subscripts on the Wilson coefficients are for the type of Lorentz structure: vector, axial-vector, scalar, 
pseudo-scalar, and tensor. The Wilson coefficients would in general be different for different fermions $f$.
Note that the Lagrangian Eq.~(\ref{L_eff1}) also contains terms that do not follow from the 
dimension six in the standard model effective field theory (SMEFT), but could be generated by 
higher order operators. This is taken into account by introducing mass and $G_F$ factors emulating 
such suppression \cite{Celis:2014asa,Hazard:2017udp}.

The last term in Eq.~(\ref{LeptNumbLagr}), ${\cal L}_{\rm eff}^{\Delta L_\mu=2}$, represents the effective 
operators changing the lepton quantum number by two units. The leading contribution to muonium 
oscillations is given by the dimension six operators.
The most general effective Lagrangian 
\begin{equation}\label{DL2}
{\cal L}_{\rm eff}^{\Delta L_\mu=2}=-\frac{1}{\Lambda^{2}}\underset{i}{\sum}C^{\Delta L=2}_{i}(\mu)Q_{i}(\mu).
\end{equation}
can be written with the operators written entirely in terms of the muon and electron degrees of freedom,
\begin{eqnarray}\label{Dim6_Op}
Q_1 &=& \left(\overline\mu_L \gamma_\alpha e_L \right)  \left(\overline\mu_L \gamma^\alpha e_L \right),  \quad
Q_2 = \left(\overline\mu_R \gamma_\alpha e_R \right)  \left(\overline\mu_R \gamma^\alpha e_R \right),
\nonumber \\
Q_3 &=& \left(\overline\mu_L \gamma_\alpha e_L \right)  \left(\overline\mu_R \gamma^\alpha e_R \right),  \quad	
Q_4 = \left(\overline\mu_L e_R \right)  \left(\overline\mu_L e_R \right), 
\nonumber \\
Q_5 &=& \left(\overline\mu_R e_L \right)  \left(\overline\mu_R e_L \right).
\end{eqnarray}
We did not include operators that could be related to the presented ones via Fierz relations. It is important to note 
that some of the operators in Eq.~(\ref{Dim6_Op}) are not invariant under the SM gauge group $SU(2)_L\times U(1)$. This 
means that they receive additional suppression, as they may be generated from the higher-dimensional 
operators in SMEFT \cite{Petrov:2016azi}.

Other $\Delta L_\mu=2$ local operators that will be important later in this paper can be written as 
\beq\label{Dim6_Op_nu}
Q_6 = \left(\overline\mu_L \gamma_\alpha e_L \right)  \left(\overline{\nu_{\mu}}_L \gamma^\alpha {\nu_e}_L \right),  \quad
Q_7 = \left(\overline\mu_R \gamma_\alpha e_R \right)   \left(\overline{\nu_{\mu}}_L \gamma^\alpha {\nu_e}_L \right),
\eeq
where we only included SMEFT operators that contain left-handed neutrinos \cite{Petrov:2016azi,Grossman:2003rw}. 
In order to see how these operators (and thus new physics) contribute to the mixing parameters, it is instructive to consider off-diagonal 
terms in the mass matrix \cite{Donoghue}
\begin{equation}\label{OffDiagonal}
\left(m-\frac{i}{2} \Gamma\right)_{12}=\frac{1}{2 M_{M}}\left\langle\ammu\left|{\cal H}_{\rm eff} \right| \mmu\right\rangle
+\frac{1}{2 M_{M}} \sum_{n} \frac{\left\langle\ammu
\left|{\cal H}_{\rm eff} \right| n\right\rangle\left\langle n\left|{\cal H}_{\rm eff} \right| \mmu\right\rangle}{M_{M}-E_{n}+i \epsilon},
\end{equation}
where the first term does not contain imaginary part, so it contributes to $m_{12}$, i.e. the mass difference. 
The second term contains bi-local contributions connected by physical intermediate states. This term 
has both real and imaginary parts and thus contributes to both $m_{12}$ and $\Gamma_{12}$.

\subsection{Mass difference: $\Delta L_\mu = 2$ operators}\label{DeltaMuTwo}

We can rewrite Eq.(\ref{OffDiagonal}) to extract the physical mixing parameters $x$ and $y$ of Eq.~(\ref{XandY}). 
For the mass difference,
\beq\label{MEforX}
x = \frac{1}{2 M_M \Gamma} \mbox{Re} \left[ 
2 \langle\ammu\left|{\cal H}_{\rm eff} \right| \mmu\rangle 
+ \langle\ammu  \left| i\int d^4x \ \mbox{T} \left[
  {\cal H}_{\rm eff} (x)   {\cal H}_{\rm eff} (0) \right]  \right | \mmu\rangle  
\right]
\eeq
Assuming the LFV NP is present, the dominant local contribution to $x$ comes from the last term in 
Eq.~(\ref{LeptNumbLagr}), 
\beq
\langle \ammu | {\cal H}_{\rm eff} |\mmu\rangle = \langle \ammu | {\cal H}_{\rm eff}^{\Delta L_\mu=2} |\mmu\rangle
\eeq
provided that only $Q_1-Q_5$ operators are taken into account. It is easy to see that the relevant contributions are 
only suppressed by $\Lambda^2$. Other contributions, including the non-local double insertions of 
${\cal L}_{\rm eff}^{\Delta L_{\mu}=1}$, represented by the second term in Eq.~(\ref{MEforX}), do contribute to the 
mass difference, but are naively suppressed by $\Lambda^4$. Thus, we shall not consider them in this paper.

In order to evaluate the mass difference contribution, we need to take the matrix elements. 
As explained in the Introduction, we expect that both spin-0 singlet and spin-1 triplet muonium states
would undergo oscillations. The oscillation parameters would in general be different, as the matrix elements 
would differ for those two cases. 

Using factorization approach familiar from the meson flavor oscillation, the matrix elements can be easily 
written in terms of the muonium decay constant $f_{M}$ \cite{Hazard:2016fnc,Fael:2018}. 
\begin{eqnarray}\label{DecayConst}
\bra{0} \overline{\mu} \gamma^{\alpha} \gamma^{5} e \ket{\mmup} &=& i f_P p^{\alpha}, \quad
\bra{0} \overline{\mu} \gamma^\alpha e \ket{\mmuv} = f_V M_{M} \epsilon^\alpha (p), 
\nonumber \\
\bra{0} \overline{\mu} \sigma^{\alpha\beta} e \ket{\mmuv} &=& i f_T \left(\epsilon^\alpha p^\beta - \epsilon^\beta p^\alpha\right),
\end{eqnarray}
where $p^\alpha$ is para-muonium's four-momentum, and $\epsilon^\alpha (p)$ is the ortho-muonium's polarization vector.
Note that $f_P=f_V=f_T=f_M$ in the non-relativistic limit. The decay constant can be written in terms of the bound-state wave function,
\begin{eqnarray}\label{Definition_fM}
f_M^2 = 4\frac{\left|\varphi(0)\right|^2}{M_{M}}, 
\end{eqnarray}
which is the QED's version of Van Royen-Weisskopf formula. For a Coulombic bound state the wave function of 
the ground state is 
\beq
\varphi (r)=\frac{1}{\sqrt{\pi a_{\tiny \mmu}^{3}}}e^{-\frac{r}{a_{\tiny \mmu}}}, 
\eeq
where $a_{\tiny \mmu}=(\alpha m_{\rm red})^{-1}$ is the muonium Bohr radius, $\alpha$ is the fine structure constant, 
and $m_{red}=m_{e}m_{\mu}/(m_{e}+m_{\mu})$ is the reduced mass. Then,
\begin{equation}\label{WFat0}
|\varphi(0)|^{2}=\frac{(m_{red}\alpha)^{3}}{\pi} =\frac{1}{\pi}  (m_{red}\alpha)^{3}.
\end{equation} 
In the non-relativistic limit factorization gives the exact result for the QED matrix elements of the six-fermion operators. 
Nevertheless, we explicitly verified that this is indeed the case (see Appendix \ref{Appendix}).  

{\bf Para-muonium}. The matrix elements of the spin-singlet states can be obtained from Eq.~(\ref{Dim6_Op})
using the definitions of Eq.~(\ref{DecayConst}),
\begin{eqnarray}\label{ME0}
 \bra{\ammup} Q_{1}\ket{\mmup} &=& \ \ f_M^2 M_M^2, \ \quad
 \bra{\ammup} Q_{2}\ket{\mmup} = \ \ f_M^2 M_M^2,  \nonumber \\ 
 \bra{\ammup} Q_{3}\ket{\mmup} &=& -\frac{3}{2} f_M^2 M_M^2, \quad
 \bra{\ammup} Q_{4}\ket{\mmup} = - \frac{1}{4}  f_M^2 M_M^2,  \nonumber \\
 \bra{\ammup} Q_{5}\ket{\mmup} &=& - \frac{1}{4}  f_M^2 M_M^2.
\end{eqnarray}
Combining the contributions from the different operators and using the definitions from Eqs.~(\ref{Definition_fM}) and (\ref{WFat0}), 
we obtain an expression for $x_P$ for the para-muonium state,
\begin{equation}\label{Delta_m_para}
x_P=\frac{4 (m_{red}\alpha)^{3}}{\pi\Lambda^2 \Gamma} 
\left[
C_1^{\Delta L= 2} + C_2^{\Delta L= 2} - \frac{3}{2} C_3^{\Delta L= 2} - 
\frac{1}{4}\left(C_4^{\Delta L= 2} + C_5^{\Delta L= 2}\right)
\right] .
\end{equation}
This result is universal and holds true for any new physics model that can be matched into a set of local 
$\Delta L = 2$ interactions. 

{\bf Ortho-muonium}. Using the same procedure, but computing the relevant matrix elements for the vector ortho-muonium state, we obtain the 
matrix elements
\begin{eqnarray}\label{ME1}
 \bra{\ammuv} Q_{1}\ket{\mmuv} &=& -3 f_M^2 M_M^2, \quad
 \bra{\ammuv} Q_{2}\ket{\mmuv} =  -3 f_M^2 M_M^2,  \nonumber \\ 
 \bra{\ammuv} Q_{3}\ket{\mmuv} &=& -\frac{3}{2} f_M^2 M_M^2, \quad
 \bra{\ammuv} Q_{4}\ket{\mmuv} = -\frac{3}{4}  f_M^2 M_M^2,  \nonumber \\
 \bra{\ammuv} Q_{5}\ket{\mmuv} &=& -\frac{3}{4}  f_M^2 M_M^2.
\end{eqnarray}
Again, combining the contributions from the different operators, we obtain an expression for $x_V$ for the ortho-muonium state,
\begin{equation}\label{Delta_m_ortho}
x_V=-\frac{12 (m_{red}\alpha)^{3}}{\pi\Lambda^2 \Gamma}
\left[
C_1^{\Delta L= 2} + C_2^{\Delta L= 2} + \frac{1}{2} C_3^{\Delta L= 2} + 
\frac{1}{4}\left(C_4^{\Delta L= 2} + C_5^{\Delta L= 2}\right)
\right].
\end{equation}
Again, this result is universal and holds true for any new physics model that can be matched into a set of local 
$\Delta L = 2$ interactions. 

It might be instructive to present an example of a BSM model that can be matched into the effective Lagrangian
of Eq.~(\ref{DL2}) and can be constrained from Eqs.~(\ref{ME0},\ref{ME1}). Let us consider a model which contains a
doubly-charged Higgs boson \cite{Swartz:1989qz,Chang:1989uk,Crivellin:2018ahj}. Such states often appear in the context of
left-right models \cite{Kiers:2005gh,Kiers:2005vx}. A coupling of the doubly charged Higgs field $\Delta^{--}$ to the lepton fields 
can be written as
\beq
{\cal L}_R =  g_{\ell \ell} \overline \ell_R \ell^c \Delta + H.c.,
\eeq
where $\ell^c=C\overline{\ell}^T$ is the charge-conjugated lepton state. Integrating out the $\Delta^{--}$ field,
this Lagrangian leads to the following effective Hamiltonian  \cite{Swartz:1989qz,Kiers:2005vx}
\beq\label{DoublyChargedHiggs}
{\cal H}_\Delta =  \frac{g_{ee} g_{\mu\mu}}{2 M_\Delta^2}  \left(\overline\mu_R \gamma_\alpha e_R \right)  
\left(\overline\mu_R \gamma^\alpha e_R \right) + H.c.,
\eeq
below the scales associated with the doubly-charged Higgs field's mass $M_\Delta$. Examining 
Eq.~(\ref{DoublyChargedHiggs}) we see that this Hamiltonian matches onto our operator $Q_2$ 
(see Eq.~(\ref{Dim6_Op})) with the scale $\Lambda = M_\Delta$ and the corresponding Wilson coefficient
$C_2^{\Delta L= 2} =  g_{ee} g_{\mu\mu}/2$.

\subsection{Width difference: $\Delta L_\mu = 2$ and $\Delta L_\mu = 1$ operators}\label{DeltaMuOne}

The lifetime difference in the muonium system can be obtained from Eq.~(\ref{OffDiagonal}) \cite{Golowich:2006gq}.
It comes from the physical intermediate states, which is signified by the imaginary part in Eq.~(\ref{OffDiagonal}) and reads,
\beq\label{YPheno}
y = \frac{1}{\Gamma} \sum_{n} \rho_n \left \langle\ammu
\left|{\cal H}_{\rm eff} \right| n\right\rangle\left\langle n\left|{\cal H}_{\rm eff} \right| \mmu\right\rangle,
\eeq
where $\rho_n$ is a phase space function that corresponds to the intermediate state that is common for 
$\mmu$ and $\ammu$. There are only two\footnote{A possible $\gamma\gamma$ intermediate state is generated by
higher-dimensional operators and therefore is further suppressed by either powers of $\Lambda$ or the QED 
coupling $\alpha$ than the contributions considered here.} possible intermediate states that can contribute to $y$,
$e^+e^-$ and $\nu\bar\nu$. The $e^+e^-$ intermediate state corresponds to a $\Delta L_{\mu}=1$ decay
$\mmu \to e^+e^-$, which implies that ${\cal H}_{\rm eff} = {\cal H}_{\rm eff}^{\Delta L_{\mu}=1}$ in
Eq.~(\ref{YPheno}). According to Eq.~(\ref{L_eff1}), it appears that, quite generally, this contribution is 
suppressed by $\Lambda^4$, i.e. will be much smaller than $x$, irrespective of the values of the 
corresponding Wilson coefficients. 

Another contribution comes from the $\nu\bar\nu$ intermediate state. This common intermediate 
state can be reached by the standard model tree level decay $\mmu \to \overline{\nu_\mu} {\nu_e}$
interfering with the $\Delta L_{\mu}=2$ decay $\ammu \to \overline{\nu_\mu} \nu_e$. Such contribution
is only suppressed by $\Lambda^2 M_W^2$ and represents the parametrically leading
contribution to $y$. We shall compute this contribution below.

Writing $y$ similarly to $x$ in Eq.~(\ref{MEforX}), i.e. in terms of the correlation function, we obtain
\beqa\label{MEforY}
y &=& \frac{1}{2 M_M \Gamma} \mbox{Im} \left[ 
\langle\ammu  \left| i\int d^4x \ \mbox{T} \left[
  {\cal H}_{\rm eff} (x)   {\cal H}_{\rm eff} (0) \right]  \right | \mmu\rangle  
\right]
\nonumber \\
&=& \frac{1}{M_M \Gamma} \mbox{Im} \left[ 
\langle\ammu  \left| i\int d^4x \ \mbox{T} \left[
  {\cal H}_{\rm eff}^{\Delta L_\mu=2} (x)   {\cal H}_{\rm eff}^{\Delta L_\mu=0} (0) \right]  \right | \mmu\rangle
\right],
\eeqa
where the ${\cal H}_{\rm eff}^{\Delta L_\mu=0}=-{\cal L}_{\rm eff}^{\Delta L_\mu=0}$ is given by the ordinary standard model Lagrangian,
\beq\label{SMLagr}
{\cal L}_{\rm eff}^{\Delta L_\mu=0} = -\frac{4 G_F}{\sqrt{2}} 
\left(\overline\mu_L \gamma_\alpha e_L \right)  \left(\overline{\nu_e}_L \gamma^\alpha {\nu_\mu}_L \right),
\eeq
and ${\cal H}_{\rm eff}^{\Delta L_\mu=2}$ only contributes through the operators $Q_6$ and $Q_7$. 
\begin{center}
\begin{figure}
\center
\includegraphics[scale=1.0]{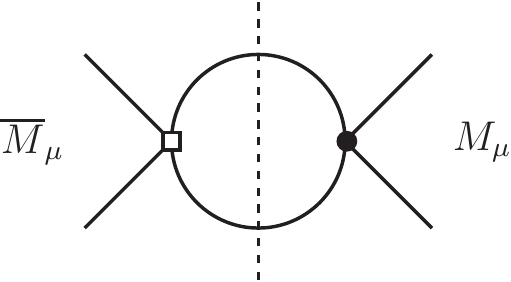}
\caption{A contribution to $y$ described in Eq.~(\ref{MEforY}). A white square represents a vertex given by Eq.~(\ref{Dim6_Op_nu}),
while a black dot is given by the SM contribution of Eq.~(\ref{SMLagr}).  A dotted line represents the imaginary part. \label{Fig:DelGam}}
\end{figure}
\end{center}

Since the decaying muon injects a large momentum into the two-neutrino intermediate state, the integral in Eq.~(\ref{MEforY})
is dominated by small distance contributions, compared to the scale set by $1/m_\mu$. We can the compute the correlation 
function in Eq.~(\ref{MEforY}) by employing a short distance operator product expansion, systematically expanding it in 
powers of $1/m_\mu$. 
\beqa\label{CorFuncY}
T &=& i\int d^4x \ \mbox{T} \left[
  {\cal H}_{\rm eff}^{\Delta L_\mu=2} (x)   {\cal H}_{\rm eff}^{\Delta L_\mu=0} (0) \right]
\nonumber \\
&=&
i\int d^4x \ \mbox{T} \left[ 
\left(\overline\mu \Gamma_\alpha e \right)  \left(\overline{\nu_{\mu}}_L \gamma^\alpha {\nu_e}_L \right) (x)
\left(\overline\mu \gamma_\beta P_L e \right) \left(\overline{\nu_e}_L \gamma^\beta {\nu_\mu}_L \right) (0) \right],
\eeqa
The leading term is obtained by contracting the neutrino fields in Eq.~(\ref{CorFuncY}) into propagators,
\beqa
\contraction{}{\overline{\nu_{\mu}}}{(x)}{\nu_\mu}
\overline\nu_{\mu}(x)\nu_\mu(0) &=& i S_F(-x),
\nonumber \\
\bcontraction{}{\nu}{_e(x)}{\overline{\nu_e}}
\nu_e(x) \ \overline{\nu_e}(0)  &=& i S_F(x),
\eeqa
where $S_F(x)$ represents the propagator in coordinate representation. In what follows we will 
consider neutrinos to be Dirac fields for simplicity. 

Using Cutkoski rules to compute the discontinuity (imaginary part) of $T$ and calculating the phase space integrals
we get 
\beq\label{Disc}
\mbox{Disc} \ T = \frac{G_F}{\sqrt{2} \Lambda^2} \frac{M_M^2}{3\pi} 
\left[C_6^{\Delta L = 2} \left(Q_1+Q_5\right) + \frac{1}{2} C_7^{\Delta L = 2} Q_3\right].
\eeq
We can now compute the lifetime difference $y$ by using Eq.~(\ref{MEforY}) and take the relevant matrix 
elements for the spin singlet and the spin triplet states of the muonium. 

{\bf Para-muonium}. The matrix elements of the spin-singlet state have been computed above and
presented in Eq.~(\ref{ME0}). Computing the matrix elements in Eq.~(\ref{MEforY}) using their definitions 
from Eqs.~(\ref{Definition_fM}) and (\ref{WFat0}), we obtain an expression for the lifetime difference $y_P$ 
for the para-muonium state,
\begin{equation}\label{Delta_Gamma_para}
y_P= \frac{G_F}{\sqrt{2} \Lambda^2} \frac{M_M^2}{\pi^2 \Gamma} (m_{red}\alpha)^{3} 
\left(C_6^{\Delta L= 2} - C_7^{\Delta L= 2}\right).
\end{equation}
It is interesting to note that if $C_6^{\Delta L= 2} = C_7^{\Delta L= 2}$ current conservation
assures that no lifetime difference is generated at this order in $1/\Lambda$ for the para-muonium.

{\bf Ortho-muonium}. Similarly, using the matrix elements for the spin-triplet state computed in Eq.~(\ref{ME1}),
the expression fo Eq.~(\ref{Disc}) leads to the lifetime difference
\begin{equation}\label{Delta_Gamma_ortho}
y_V= -\frac{G_F}{\sqrt{2} \Lambda^2} \frac{M_M^2}{\pi^2 \Gamma} (m_{red}\alpha)^{3} 
\left( 5 C_6^{\Delta L= 2} + C_7^{\Delta L= 2}\right),
\end{equation}
We emphasize that Eqs.~(\ref{Delta_Gamma_para}) and (\ref{Delta_Gamma_ortho}) represent  
parametrically leading contributions to muonium lifetime difference, as they are only suppressed by 
two powers of $\Lambda$. 

\section{Experimental constraints}\label{Constraints}

We can now use the derived expressions for $x$ and $y$ to place constraints on the BSM 
scale $\Lambda$ (or the Wilson coefficients $C_i$) from the experimental constraints on 
muonium-anti-muoium oscillation parameters. Since both spin-0 and spin-1 muonium states 
were produced in the experiment \cite{Willmann:1998gd}, we should average the oscillation 
probability over the number of polarization degres of freedom, 
\beq
P(\mmu \rightarrow \ammu)_{\rm exp} = \sum_{i=P,V} \frac{1}{2S_i+1} P(\mmu^i \rightarrow \ammu^i),
\eeq
where $P(\mmu \rightarrow \ammu)_{\rm exp}$ is the experimental oscillation probability 
from Eq.~(\ref{Exp_constr}). We shall use the values of $S_B(B_0)$ for 
$B_0 = 2.8$ $\mu$T from the Table II of \cite{Willmann:1998gd}, as it will provide us the 
best experimental constraints on the BSM scale $\Lambda$.
We report those constraints in Table \ref{Table1}.
\begin{table}[t]
\begin{center}
\begin{tabular}{|c|c|c|c|}
\hline\hline
$~$Operator $~$ & $~~$ Interaction type $~~$ & $~S_B(B_0)$ (from \cite{Willmann:1998gd}) $~~$ & $~~$Constraints on the scale $\Lambda$, TeV $~~$ \\
\hline\hline
$Q_1$ & $(V-A)\times (V-A)$ & 0.75 & $5.4$ \\
$Q_2$ & $(V+A)\times (V+A)$ & 0.75 & $5.4$ \\
$Q_3$ & $(V-A)\times (V+A)$ & 0.95 & $5.4$ \\
$Q_4$ & $(S+P)\times (S+P)$ & 0.75 & $2.7$ \\
$Q_5$ & $(S-P)\times (S-P)$ & 0.75 & $2.7$ \\
$Q_6$ & $(V-A)\times (V-A)$ & 0.75 & $0.58\times 10^{-3}$ \\
$Q_7$ & $(V+A)\times (V-A)$ & 0.95 & $0.38\times 10^{-3}$ \\
\hline
\hline
\end{tabular}
\end{center}
\caption{Constraints on the energy scales probed by different $\Delta L=2$ operators of Eqs.~(\ref{Dim6_Op}) and (\ref{Dim6_Op_nu}). 
We set the corresponding Wilson coefficient $C_i=1$.}
\label{Table1}
 \end{table}
As one can see from Eqs.~(\ref{Delta_m_para}), (\ref{Delta_m_ortho}), (\ref{Delta_Gamma_para}), and (\ref{Delta_Gamma_ortho}),
each observable depends on the combination of the operators. We shall assume that only one operator at a time gives a dominant 
contribution. This ansatz is usually referred to as the single operator dominance hypothesis. It is not necessarily realized in many
particular UV completions of the LFV EFTs, as cancellations among contributions of different operators are possible. It is 
however a useful tool in constraining parameters of ${\cal L}_{\rm eff}$.

Since it is the combination $C_i/\Lambda^2$ that enters the theoretical predictions for $x$ and $y$, one cannot separately measure $C_i$ 
and $\Lambda$. We choose to constrain the scale $\Lambda$ that is probed by the corresponding operator and set the corresponding 
value of the Wilson coefficient $C_i$ to one. Such approach, as any calculation based on effective field-theoretic techniques, 
has its advantages and disadvantages. The advantage of such approach is in the fact that it allows to constrain {\it all} possible models of 
New Physics that can generate $\mmbar$ mixing. The models are encoded in the analytic expressions for the Wilson coefficients 
of a few effective operators in Eq.~(\ref{Dim6_Op}). The disadvantage is reflected in the fact that possible complementary studies of 
New Physics contributions to $\Delta L = 1$ and $\Delta L = 2$ processes are not straightforward. Those can be done by considering particular 
BSM scenarios, which is beyond the scope of this paper\footnote{An example of such analysis concentrating on models 
containing doubly-charged Higgs states is \cite{Crivellin:2018ahj}, where it was concluded that 1999 data on
 $\mmbar$ oscillations \cite{Willmann:1998gd} give constraints that are weaker than (but complimentary to) those obtained from a 
 combination of constraints on $\mu \to 3e$ and other experiments. Other examples include models where the mixing is generated by 
 loops with neutral particles, such as heavy neutrinos}. The EFT techniques are then used to simplify calculations of radiative 
 corrections. 

The results are reported in Table~\ref{Table1}. As can be seen, the experimental data provide constraints on the scales comparable to 
those probed by the LHC program, except for $Q_6$ and $Q_7$. The results indicate that existing bounds on $\mmbar$
oscillation parameters probe NP scales of the order of several TeV. The constraints on the lepton-flavor violating neutrino operators 
$Q_6$ and $Q_7$ are understandably weaker, as the lifetime difference is suppressed by a factor $G_F/\Lambda^2$, while
the mass difference is only suppressed by a factor of $1/\Lambda^2$.
We would like to emphasize that constraints on the oscillation parameters come from the data that is over 20 years 
old \cite{Willmann:1998gd}! We find it amazing that the data obtained over two decades ago probe the same energy scales 
as current LHC experiments. 

We urge our experimental colleagues to further study muonium-antimuonium oscillations. It would be interesting to see how far 
the proposed MACE experiment \cite{MACE} or similar facility at FNAL could push the constraints on the muonium oscillation 
parameters.

\section{Conclusions}\label{Conclusions}

Lepton flavor violating transitions provide a powerful engine for new physics searches. In this work we 
revisited phenomenology of muonium-antimuonium oscillations. We argued that in generic models of 
new physics both mass and lifetime differences in the muonium system would contribute to the 
oscillation probability. We computed the normalized mass difference $x$ in the muonium system with the 
most general set of effective operators for both spin-singlet and the spin-triplet muonium states.
We set up a formalism for computing the lifetime difference and computed the parametrically 
leading contribution to $y$. Using the derived expressions for $x$ and $y$ we then put constraints on the BSM scale $\Lambda$. 
From this we found that for operators $Q_1-Q_5$ the experimental data provided constraints on scales relevant to the LHC program.

\begin{acknowledgments}
This work was supported in part by the U.S. Department of Energy under contract de-sc0007983. AAP thanks the 
Institute for Nuclear Theory at the University of Washington for its kind hospitality and 
stimulating research environment. This research was also supported in part by the INT's 
U.S. Department of Energy grant No. DE-FG02- 00ER41132.
\end{acknowledgments}

\section{Appendix}\label{Appendix}

In this Appendix we show that the vacuum insertion approximation leads to the same answer as a direct computation of a four-fermion 
matrix element relevant for the muonium-anti-muonium oscillations. We shall show that by computing a matrix element of 
the $Q_1$ operator as an example. The matrix elements is defined as
\begin{equation}\label{Q_1_matrix_element}
\left\langle Q_1 \right\rangle = \bra{\overline{M}_{\mu}}\left(\overline\mu \gamma_\alpha P_L e \right)  \left(\overline\mu \gamma^\alpha P_L e \right)\ket{M_{\mu}}
\end{equation}
for both pseudoscalar and vector muonium states.
In order to compute the matrix element in Eq.~(\ref{Q_1_matrix_element}) we need to build the muonium states. We can employ the standard 
Bethe-Salpeter formalism. Since the muonium state is essentially a a nonrelativistic Coulomb bound state of a $\mu^+$ and an $e^-$, 
we can conventionally define it \cite{Peskin:1995ev} 
\begin{equation}\label{Mu_bound_state}
\ket{\mmu} = \sqrt{\frac{2M_{M}}{2m_\mu 2m_e}}\int\frac{d^{3}p}{(2\pi)^{3}}\widetilde\varphi(\mathbf{p})\ket{\mathbf{p},\mathbf{p}'}.
\end{equation}
This state is normalized as $\braket{M_{\mu}(\textbf{P})}{M_{\mu}(\textbf{P}')} = 2E_{\textbf{p}}(2\pi)^{3}\delta^{3}(\textbf{P}-\textbf{P}')$. The muonium
state in Eq.~(\ref{Mu_bound_state}) is projected from a two-particle state of a muon and an electron  
$\ket{p,p'} = \sqrt{2E_{\textbf{p}}}\sqrt{2E_{\textbf{p}'}} \ a^{(e)\dagger}_{\textbf{p}} b^{(\mu)\dagger}_{\textbf{p}'}\ket{0}$ with the help of 
the Fourier transform of the spatial wave equation describing the bound state $\widetilde\varphi(p)$, 
\beq
\widetilde\varphi(\mathbf{p}) = \int d^3r\varphi(\mathbf{r})e^{i\mathbf{p}\mathbf{r}} .
\eeq
We expand each electron and muon field in the operator of Eq.~(\ref{Q_1_matrix_element}) as 
\begin{equation}\label{field expansions}
\psi(x) = \int\frac{d^{3}p}{(2\pi)^{3}}\frac{1}{\sqrt{2E_{\textbf{p}}}}\sum_{s}\left(a_{\textbf{p}}^{s}u^{s}(p)e^{-ipx} + b_{\textbf{p}}^{s\dagger}v^{s}(p)e^{ipx}\right).
\end{equation}
We will work in non-relativistic approximation and neglect the momentum dependence of the spinors, which are defined as
\begin{eqnarray}\label{non relativistic spinors}
u &=& \sqrt{m_{e}}\begin{pmatrix}
\xi \\
\xi
\end{pmatrix},  \qquad \qquad
v = \sqrt{m_{e}}\begin{pmatrix}
\eta \\
-\eta 
\end{pmatrix}, 
\nonumber \\
\overline{u} &=& \sqrt{m_{\mu}}\begin{pmatrix}
\xi^{\dagger}, \xi^{\dagger}
\end{pmatrix}
\gamma^{0},  \qquad
\overline{v} = \sqrt{m_{\mu}}\begin{pmatrix}
\eta^{\dagger}, -\eta^{\dagger} 
\end{pmatrix}
\gamma^{0}.
\end{eqnarray}  
Here $\xi$ and $\eta$ are the two-component spinors \cite{Peskin:1995ev}. There are four ways to Wick contract the fields 
in the operator with those generating the state. Using anti-commutation relation 
$\{a_{\textbf{p}},a^{\dagger}_{\textbf{p}'}\} = (2\pi)^{3}\delta^{3}(\textbf{p}-\textbf{p}')$ results in
\begin{eqnarray}\label{<Q_1> after wick contractions}
\left\langle Q_1 \right\rangle &= &
\left[\left(\overline{u} \gamma_\alpha P_L v \right)  \left(\overline{v} \gamma^\alpha P_L u \right) + 
\left(\overline{v} \gamma_\alpha P_L u \right)  \left(\overline{u} \gamma^\alpha P_L v \right) \right. 
\nonumber \\
&-& \left. \left(\overline{v} \gamma_\alpha P_L v \right)  \left(\overline{u} \gamma^\alpha P_L u \right) - 
\left(\overline{u} \gamma_\alpha P_L u \right)  \left(\overline{v} \gamma^\alpha P_L v \right)\right]_{\mmu}
\times \left|\int\frac{d^{3}p}{(2\pi)^{3}}\widetilde\varphi(\bf{p})\right|^{2},
\end{eqnarray}
where we indicated that the spinors still need to be projected onto the spin-triplet or or the spin-singlet states.
This projection can be illustrated explicitly by considering the first term in Eq.~(\ref{<Q_1> after wick contractions}), 
$\left(\overline{u} \gamma_\alpha P_L v \right)  \left(\overline{v} \gamma^\alpha P_L u \right)$, the rest can be 
computed in a complete analogy to that. Employing the Weyl basis for the gamma matrices,
\begin{eqnarray}\label{gamma matrices}
\gamma^{0} = 
\begin{pmatrix}
0 & 1 \\ 
1 & 0
\end{pmatrix},  \qquad
\gamma^{\alpha} = 
\begin{pmatrix}
0 & \sigma^{\alpha} \\ 
\overline{\sigma}^{\alpha} & 0
\end{pmatrix},  \qquad
\gamma^{5} = 
\begin{pmatrix}
-1 & 0 \\ 
0 & 1
\end{pmatrix},
\end{eqnarray}
where $\sigma^{\alpha}$ and $\overline{\sigma}^{\alpha}$ are defined as
\begin{equation}\label{sigma four vectors}
\sigma^{\alpha} = \left(\mathbf{1}, \vec{\sigma} \right), 
 \qquad
\overline{\sigma}^{\alpha} = \left( \mathbf{1}, -\vec{\sigma} \right).
\end{equation}
Note that $\vec{\sigma}$ is a vector comprised of the Pauli matrices, and 
$\mathbf{1}$ is the 2 $\times$ 2 identity matrix. 
Now, expanding the matrix elements,
\begin{align}\label{operator expansion}
\left(\overline{u} \gamma_\alpha P_L v \right)  \left(\overline{v} \gamma^\alpha P_L u \right)_{\mmu} = &\frac{1}{4}m_{\mu}m_e
\begin{pmatrix}
\xi^{\dagger}, & \xi^{\dagger}
\end{pmatrix}
\gamma^{0}\gamma^{\alpha}\left(1-\gamma^{5}\right)
\begin{pmatrix}
\eta\\
-\eta
\end{pmatrix} \nonumber \\
\times & 
\begin{pmatrix}
\eta^{\dagger}, & -\eta^{\dagger}
\end{pmatrix}
\gamma^{0}\gamma_{\alpha}\left(1-\gamma^{5}\right)
\begin{pmatrix}
\xi\\
\xi
\end{pmatrix}, 
\end{align}
or writing out the gamma matrices and spinors from Eqs.~(\ref{gamma matrices}) and (\ref{sigma four vectors}) and
making rearrangements we find
\begin{eqnarray}\label{operator expansion w/ matrices}
\left(\overline{u} \gamma_\alpha P_L v \right)  \left(\overline{v} \gamma^\alpha P_L u \right)_{\mmu} 
&=& m_{\mu}m_e
\begin{pmatrix}
\xi^{\dagger}, & \xi^{\dagger}
\end{pmatrix}
\begin{pmatrix}
0 & 1 \\ 
1 & 0
\end{pmatrix}
\begin{pmatrix}
0 & \sigma^{\alpha} \\ 
\overline{\sigma}^{\alpha} & 0
\end{pmatrix}
\begin{pmatrix}
1 & 0 \\ 
0 & 0
\end{pmatrix}
\begin{pmatrix}
\eta\\
-\eta
\end{pmatrix} 
\nonumber \\
& \times & 
\begin{pmatrix}
\eta^{\dagger}, & -\eta^{\dagger}
\end{pmatrix}
\begin{pmatrix}
0 & 1 \\ 
1 & 0
\end{pmatrix}
\begin{pmatrix}
0 & \sigma_{\alpha} \\ 
\overline{\sigma}_{\alpha} & 0
\end{pmatrix}
\begin{pmatrix}
1 & 0 \\ 
0 & 0
\end{pmatrix}
\begin{pmatrix}
\xi\\
\xi
\end{pmatrix} 
\nonumber \\
&=& m_\mu m_e \left(\xi^{\dagger} \overline{\sigma}^{\alpha} \eta \right)
\left(\eta^{\dagger} \overline{\sigma}_{\alpha} \xi \right)_{\mmu} 
\nonumber \\
&=& m_\mu m_e \text{Tr}\left[\eta \xi^{\dagger} \overline{\sigma}^{\alpha} \right]
\text{Tr}\left[\xi \eta^{\dagger} \overline{\sigma}_{\alpha} \right]_{\mmu} .
\end{eqnarray}
Projection onto the singlet (spin-0) or the triplet (spin-1) states can be achieved through the 
substitutions \cite{Peskin:1995ev},
\begin{equation}\label{spin-0 spinor product}
\xi\eta^{\dagger} = \frac{1}{\sqrt{2}}\mathbf{1}_{2\times2}
\end{equation}
for the spin-0 state and 
\begin{equation}\label{spin-1 spinor product}
\xi\eta^{\dagger} = \frac{1}{\sqrt{2}}\vec {\epsilon}^{~*}\cdot \vec {\sigma}
\end{equation}
for the spin-1 state with three possible polarization states, $\vec{\epsilon}_1 = (0,0,1)$, $\vec{\epsilon}_2 = \frac{1}{\sqrt{2}}(1,i,0)$, 
and $\vec{\epsilon}_3 = \frac{1}{\sqrt{2}}(1,-i,0)$. It is convenient to introduce polarization four-vectors \cite{Fael:2018},
$\epsilon_{\nu}^{*}= (0,\vec{\epsilon^{*}})$, $\sigma^{\nu}=(\mathbf{1},\vec{\sigma})$, and $\overline{\sigma}^{\nu}=(\mathbf{1},-\vec{\sigma})$.  

Computing the traces for the singlet spin state, Eq.~(\ref{operator expansion w/ matrices}) becomes
\begin{equation}\label{operator expansion trace spin-0}
m_\mu m_e \text{Tr}\left[\eta \xi^{\dagger} \overline{\sigma}^{\alpha} \right]_{\mmup} 
\text{Tr}\left[\xi \eta^{\dagger} \overline{\sigma}_{\alpha} \right]_{\mmup}  
= \frac{1}{2}\text{Tr}[\overline{\sigma}^{\alpha}]\text{Tr}[\overline{\sigma}_{\alpha}] = 2 m_{\mu} m_e.
\end{equation}
Notice that this expression is zero unless $\alpha = 0$. Similarly, for the spin-1 state Eq.~(\ref{operator expansion w/ matrices}) becomes
\begin{eqnarray}\label{operator expansion trace spin-1}
m_\mu m_e \text{Tr}\left[\eta \xi^{\dagger} \overline{\sigma}^{\alpha} \right]_{\mmuv} 
\text{Tr}\left[\xi \eta^{\dagger} \overline{\sigma}_{\alpha} \right]_{\mmuv}  
&=& \frac{1}{2}m_\mu m_e\epsilon_{\mu}\epsilon_{\nu}^{*} \ \text{Tr}[\overline{\sigma}^{\alpha}\sigma^{\mu}]\text{Tr}[\overline{\sigma}_{\alpha}\sigma^{\nu}]
\nonumber \\
&=& 2 m_{\mu} m_e\epsilon_{\mu}\epsilon^{\mu*}
= -6 m_{\mu}m_e,
\end{eqnarray}
as the sum over polarizations is $\epsilon_{\mu}\epsilon^{\mu*} = -3$. Following the same procedure for the rest of the terms in Eq. (\ref{<Q_1> after wick contractions}) 
and using 
\begin{equation}\label{spacial wavefuntion r=0}
\left|\int\frac{d^{3}p}{(2\pi)^{3}}\widetilde\varphi(p)\right|^{2} = |\varphi(0)|^{2},
\end{equation}
we get $\left\langle Q_1 \right\rangle$ for spin-0 and spin-1
\begin{eqnarray}
\bra{\ammup}  Q_1 \ket{\mmup} = 4M_{M} |\varphi(0)|^{2}, \qquad
\bra{\ammuv}  Q_1 \ket{\mmuv} = -12M_{M} |\varphi(0)|^{2},
\end{eqnarray}
which is identical to the definitions in Eq.~(\ref{ME0}) and (\ref{ME1}), provided that 
the Van Royen-Weisskopf formula of Eq.~(\ref{Definition_fM}) is used. 
The proof for the rest of the operators follows the same steps.


\end{document}